\documentclass[fleqn,12pt,twoside]{article}
\usepackage{espcrc1}

\usepackage[figuresright]{rotating}

\newcommand{\AmS}{{\protect\the\textfont2
  A\kern-.1667em\lower.5ex\hbox{M}\kern-.125emS}}

\def\be{\begin{equation}}
\def\ee{\end{equation}}

\title{Microscopic Model of Charge Carrier Transfer in Complex Media}

\author{O.B\'enichou\address[LPTMC]{Laboratoire de Physique Th\'eorique de la Mati\`ere Condens\'ee,\\
 Universit\'e Paris 6,
4, Place Jussieu, 75252 Paris, France},
         J.Klafter\address{School of Chemistry,\\ Tel Aviv University,
Tel Aviv 69978, Israel},
        M.Moreau\addressmark[LPTMC],
        and
        G.Oshanin\addressmark[LPTMC]\address{Department of
Inhomogeneous Condensed Matter Theory, Max-Planck-Institut f\"ur
Metallforschung,  Heisenbergstrasse 3, D-70569 Stuttgart,
Germany}\thanks{Corresponding author. e-mail:
oshanin@lptl.jussieu.fr.}}

\begin{document}

\maketitle

\begin{abstract}
We present a microscopic model of a charge carrier transfer under an
action of a constant electric field in a complex medium.
Generalizing previous theoretical approaches, we model the dynamical
environment hindering the carrier motion by dynamic percolation,
i.e., as a medium comprising particles which move randomly on a
simple cubic lattice, constrained by hard-core exclusion, and may
spontaneously annihilate and re-appear  at some prescribed rates. We
determine analytically the density profiles of the "environment"
particles, as seen from the stationary moving charge carrier,  and
calculate its terminal velocity, $V_{c}$,  as the function of the
applied field and other system parameters. We realize that for
sufficiently small external fields the force exerted  on the carrier
by the "environment"
 particles shows a viscous-like behavior and
define an analog of the Stokes formula for such dynamic percolative
environments. The corresponding friction coefficient is also
derived.   We show  that
 the density profile of the environment particles is
strongly inhomogeneous: In front of the stationary
 moving charge carrier the density is higher than the average density, $\rho_s$, and approaches it as an exponential function of the distance from the carrier.
Behind the carrier the local density is lower than $\rho_s$ and
 the approach towards  $\rho_s$ may proceed differently depending on whether the particles
number is or is not explicitly conserved.
\end{abstract}

\section{INTRODUCTION}

The percolation concept has been a key notion for understanding
transport and conduction processes in a wide range of complex
disordered media. A few stray examples are ionic conduction in
polymeric, amorphous or glassy ceramic  electrolytes, diffusion in
biological tissues and permeability of disordered membranes
\cite{0,1,2,3}.

Most of physical situations studied so far (see Refs.\cite{0,1,2,3}
and references therein) concern systems in which complex disordered
environment can be considered as "frozen"; that is, the random
environment in which a given transfer process takes place does not
change in time. This is certainly the case in many instances, but it
is not true in general. In fact, there are many experimental systems
in which the static percolation picture is not valid since the
structure of the host material reorganizes itself on a time scale
comparable to that at which the transfer itself occurs. Such a
reorganization happens, namely, in certain biomembranes \cite{4},
solid protonic conductors \cite{5}, oil-continuous microemulsions
\cite{6,7,8,9} or polymer electrolytes \cite{10,11,12}.

In particular, ionic transport across a biomembrane, such as,
 e.g., gramicidin-$A$,
 proceeds by the motion of ions through
molecular channels along which they encounter potential barriers
whose heights fluctuate randomly in time. The fluctuations of
potential barriers may impede the transport
 constituting
an important controlling  factor   \cite{4}. In the case of protonic
conduction by the Grotthus mechanism \cite{5}, site-to-site hopping
of charge carriers takes place between neighboring $H_2O$ or $NH_3$
groups that have a favorable
 relative orientation. Here, the structural
host-reorganization process
 interacting with the carrier motion occurs due to
thermally activated rotation of the $H_2O$ or $NH_3$ groups. In a
similar fashion, within
 oil-continuous microemulsions,
 the charge transport
proceeds by charge being transferred from one water globule
 to another, as globules approach each other in
their Brownian motion \cite{6,7,8,9}. Last but not least, in polymer
electrolytes, such as, e.g. polyethylene oxide complexed
non-stoichiometrically with the ionic salt $NaSCN$, the $Na^+$ ions
are largely tetrahedrally coordinated by polyether oxygens, but
 at the same time that $Na^+$ ions hop from
one fourfold coordination site to another, the oxygens
 themselves, along with the polymeric
backbone, undergo large-amplitude wagging and even
 diffusive motion \cite{10,11,12}.

Clearly, all these examples involve two characteristic
 time scales: the typical
time $\tau$ between two successive hops of the charge carrier and a
typical renewal time
 $\tau^*$ of the environment itself; that is,  the time at which
 the host complex medium re-organizes itself
 sufficiently enough to provide a new
set of available pathways for charge carrier transfer. Accordingly,
the conventional static percolation picture can be strictly valid
when only the characteristic time $\tau^*$  gets infinitely large.
Only in this case one may expect an anomalous large-scale dynamics.
On contrary, when $\tau^*$  is finite, \textit{dynamic percolation}
picture has to be applied. In this situation, one encounters  quite
a different behavior when compared to the random environments with
quenched disorder. As a result,  one expects an  Ohmic-type or
Stokes-type linear velocity-force relation for the carrier's
terminal velocities as a function of the applied field, in contrast
 to the threshold behavior and anomalous dynamics
predicted by the conventional static percolation theory.

The prefactor in the
 linear velocity-force  relation may depend, however, in a non-trivial way
on the system's parameters and this  dependence constitutes  the
main challenge for the theoretical analysis here. On the other hand,
we note that in the above mentioned examples of the dynamic
percolative environments quite different physical processes are
responsible for the time evolution of the host medium. Consequently,
one expects that the prefactor in the Stokes-type velocity-force
relation should also be dependent
 on the precise mechanism which underlies the temporal
re-organization of the environment.

Theoretical modeling of charge carrier transfer in dynamic
percolative environments has followed basically two different lines
of thought. Early models
 of dynamic percolation \cite{drugerA,drugerB} described the
random environment within the framework of a standard
bond-percolation model allowing the strength of each bond to
fluctuate in time between zero and some finite value. In this
approach, the dynamics of the host medium
 \cite{drugerA,drugerB} was accounted for  by a series of instantaneous
renewal events. These events were assumed to occur at random times,
chosen from  a renewal time
 distribution. In the renewal process the
positions of all unblocked bonds are being reassigned,
 such that after each renewal
event a carrier sees a newly defined network.  This approach is thus
characterized by a \textit{global  dynamical disorder} without
global conservation laws and correlations,
 since the entire
set of random hopping rates is renewed independently of the previous
history. Another model characterized by a \textit{local  dynamical
disorder} has been proposed in Refs.\cite{zwanzig}
 and
\cite{sahimi}, and subsequently generalized to
 the non-Markovian case in Ref.\cite{chatterjee}.
The difference of this model in regard to the previous one is that
here the hopping rates at different
 sites fluctuate \textit{independently} of each other.
That is, states of individual bonds, rather than that of the whole
lattice change in the renewal events. To describe the dynamical
behavior in the \textit{local  dynamical disorder} case,   a
dynamical
 mean-field theory has been proposed \cite{zwanzig,sahimi},
based on the effective medium approximation introduced for the
 analysis of random walks on lattices with static
disorder \cite{odagaki}. Subsequently, this model has been
generalized  to include the
 possibility of multistate transformations of the
\textit{dynamically} random complex medium \cite{granekB}. More
recently,
 several exactly
solvable one-dimensional models with \textit{global} and
 \textit{local}  dynamical disorder have been discussed \cite{garcia}.

The second  approach to the problem emerged within the context of
ionic conductivity in superionic solids. Here, the dynamical
percolative environment has been considered as a multicomponent
mixture of mobile species in which one or several neutral components
block
 the carrier component   \cite{hilferA}.  In particular, such a situation can be observed in
 a superionic
conductor $\beta''$-alumina, doped with two different ionic species
(e.g. $Na^+$ and $Ba^{2+}$), where small $Na^+$ ions are rather
mobile, while the larger $Ba^{2+}$ ions move essentially slower and
temporarily block the $Na^+$ ions. Contrary to the previous line of
thought, the dynamics of such a percolative environment has
essential correlations, generated by hard-core exclusion
interactions  between the species involved. Moreover, a salient
feature of these situations is that here the total number of the
particles involved is conserved, i.e., dynamics of the environment
is globally constrained by a certain "conservation law". The impact
of this property on charge carrier transfer will become clearer as
we proceed.

Further on, in Ref.\cite{hilferA} the frequency-dependent ionic
conductivity of the light species has been analyzed combining a
continuous time random walk approach for the dynamical problem with
an effective medium approximation describing the frozen environment
of slow species. Next, as an explanation of the sharp increase of
electrical conductivity transition in water-in-oil microemulsions
when the volume fraction of water is increased towards a certain
threshold value, in Refs.\cite{7} and \cite{8} it has been proposed
that the charge carriers are not trapped in the finite water
clusters, but rather a charge on a water globule can propagate by
either hopping to a neighboring globule, when they approach each
other, or via the diffusion of the host globule itself. This picture
has been interpreted in terms of a model similar to that employed in
Ref.\cite{hilferA}, with the only difference being that here the
"blockers" of Ref.\cite{hilferA} play the role of the transient
charge carriers. In the model of Refs.\cite{7} and \cite{8}, in
which the host dynamics is influenced by spatial correlations and
conservation of the number of the water globules involved,  the
conductivity depends on the rate of cluster rearrangement. Lastly, a
similar problem of a carrier diffusion in an environment created by
mobile hard-core
 lattice-gas particles has
been analyzed in Ref.\cite{granekS} by using the developed dynamic
bond
 percolation theory  of Refs.\cite{drugerA} and
\cite{drugerB}.

In this paper, we pursue our previous analysis of the charge carrier
transfer in complex media \cite{oliv} considering a generalize
dynamic percolation model which shares common features with both
bond-fluctuating models of
Refs.\cite{drugerA,drugerB,zwanzig,sahimi,chatterjee,granekB,garcia}
and those involving mobile blockers of Refs.\cite{hilferA,granekS}.
The system we consider consists of a host lattice, which here is a
regular cubic lattice whose sites support at most a single
occupancy, hard-core "environment" particles, and a single,
hard-core charge carrier particle subject to external electric
field. The "environment" particles move on the lattice by performing
a random hopping between the neighboring lattice sites, which is
constrained by the hard-core interactions, and may disappear from
and re-appear (renewal processes) on the empty sites of the lattice
with  some prescribed rates. Contrary to the environment particles
the charge carrier can not disappear spontaneously, and is subject
to a constant external force ${\vec E}$. Hence, the carrier performs
a biased random walk, which is constrained by the  hard-core
interactions with the "environment" particles. In consequence, on
may consider the carrier as some "probe" designed to measure the
response of the percolative environment to the internal perturbation
or, in other words, the frictional properties of such a dynamical
environment.

Now, a salient feature  of our model, which makes it different to
the previously proposed models of dynamic percolation, is that here
the interaction between "environment" particles and the carrier are
included, such that the latter may influence the dynamics of the
environment. This results, as we proceed to show, in the emergence
of complicated density profiles of the "environment" particles
around the carrier. These profiles, as well as the terminal velocity
$V_{c}$ of the carrier, are determined here explicitly, in terms of
an approximate approach of Ref.\cite{burlatsky}, which is based on
the decoupling of the triple carrier-particle-particle
 correlation functions into the product of pair
correlations.

We realize that the "environment" particles tend to accumulate in
front of the driven charge carrier creating a "traffic jam", which
impedes
 its motion. The density profiles  around the carrier are
asymmetric: the local density of the "environment"
 particles in front of the carrier is higher than the
average and approaches the average value as an exponential
 function of the distance from the carrier. The characteristic
length and the amplitude of the density relaxation function are
calculated explicitly. On the other hand, behind the carrier, i.e.
in its "wake", the local density is lower than the average.  We find
that, surprisingly, the functional form of the density profile might
be very different depending on
 the condition whether the number of particles in the percolative environment
is explicitly conserved or not; as a matter of fact, the local
density behind the carrier
 may tend to the average value either as an exponential or even as an
 \textit{algebraic} function of the distance, revealing in the latter case
especially strong memory effects and strong correlations between the
particle distribution in the
 environment and the carrier position.
Further on, we find
 that the terminal velocity of the carrier particle depends
 explicitly on the excess density in the "jammed" region in
 front of the carrier, as well as on the "environment" particles density past the carrier. Both, in turn, are
dependent on the
 magnitude of the velocity,
as well as on the rate of the renewal processes and the
 rate at which the "environment" particles can diffuse away from the carrier.
 The interplay between the jamming effect of the environment, produced by the carrier
particle, and the rate of its homogenization due to diffusive
smoothening  and  renewal processes, manifests itself
 as a medium-induced frictional force exerted on the carrier, whose
magnitude depends on the carrier velocity. As a consequence of such
a non-linear coupling, in the general case, (i.e. for arbitrary
rates of the renewal and diffusive processes), $V_{c}$ can be found
only implicitly, as the solution of a non-linear equation relating
$V_{c}$ to the system parameters. This equation simplifies
considerably in the limit of small applied external fields ${\vec
E}$ and we find that the force-velocity relation to the field
becomes linear. This implies that the frictional force exerted on
the carrier particle by the environment is \textit{viscous}. This
linear force-velocity relation can be therefore
 interpreted as the analog of the Stokes
formula for the dynamic percolative environment under study; in this
case, the carrier velocity is calculated explicitly as well as the
corresponding friction coefficient. Assuming that the Einstein
relation between the carrier mobility and its diffusion coefficient
holds \cite{ein}, we then estimate the self-diffusion coefficient of
the carrier in
 absence of external field.
We show that when only diffusive re-arrangement of the percolative
environment is allowed, while the renewal processes are suppressed,
the general expression for the diffusion coefficient  reduces to the
one obtained previously in Refs.\cite{nakazato} and \cite{elliott}.
We note  that the result of Refs.\cite{nakazato} and \cite{elliott}
is known to serve as a very good approximation for the
 self-diffusion coefficient in hard-core
lattice-gases \cite{kehr}.

The paper is structured as follows: In Section
 II we formulate
the model and introduce basic notations.  In Section III we write
down the dynamical equations which govern the time evolution of the
"environment" particles and of the carrier. Section IV is devoted to
the analytical solution of these evolution equations in the limit $t
\to \infty$. In Section V we derive explicit asymptotic results for
the carrier terminal velocity in the limit of small applied external
fields ${\vec E}$ and obtain the analog of the Stokes formula for
such a percolative environment. Asymptotic behavior of the density
profiles of the "environment" particles around the carrier is
discussed in Section VI. Finally, we conclude in Section VII with a
brief summary and discussion of our results.

\section{THE MODEL.}

The model for charge carrier transfer in complex media consists of a
three-dimensional simple cubic lattice of spacing $\sigma$, the
sites of which are partially occupied by identical hard-core
"environment" particles and a single, hard-core, carrier particle.
For both types of particles the hard-core interactions prevent
multiple occupancy of the lattice sites;
 that is, no two "environment" particles or the "carrier" and an "environment" particle
 can occupy simultaneously
the same site, and particles can not pass through each other.

The occupation of the lattice sites by the "environment" particles
is characterized by the time-dependent occupation variable
$\eta({\vec r})$, ${\vec r}$ being the lattice-vector of the site in
question. This variable assumes two values:
\begin{equation}
\eta({\vec r}) = \left\{\begin{array}{ll}
1,     \mbox{ if the site ${\vec r}$ is occupied} \nonumber\\
0,     \mbox{ if the site ${\vec r}$  is empty}
\end{array}
\right.
\end{equation}
Next, the dynamics of the "environment" particles is defined via the
following rules The particles may, at a given rate, spontaneously
disappear from the lattice, and may re-appear at random positions
and random time moments, which is reminiscent of the host medium
dynamics stipulated in
Refs.\cite{drugerA,drugerB,zwanzig,sahimi,chatterjee,granekB,garcia}.
These two processes will be referred to, in general, as renewal
processes.

Now, the environment particles move randomly by performing
nearest-neighbor random walks constrained by the hard-core
interactions, which is the main feature of the approach in
Refs.\cite{hilferA,granekS}. More specifically, we stipulate that
any of the "environment" particles waits a time $\delta \tau$, which
has an exponential probability distribution with a mean $\tau^*$,
and then chooses from a few possibilities: (a) disappearing from the
lattice at rate $g/\tau^*$, which is realized instantaneously,
 or (b) attempting to hop, at rate $l/6 \tau^*$, onto one of $6$ neighboring sites.
The hop is actually fulfilled if the target site is not occupied at
this time moment by any other particle; otherwise, the particle
attempting to hop remains at its initial position, and (c) particles
may re-appear on any \textit{vacant} lattice site with rate
$f/\tau^*$.

Note that, for simplicity, we  assumed  that the characteristic
diffusion time and the renewal times of the "environment" particles
are equal to each other. These times, i.e. $\tau_{dif}$, mean
creation time $\tau_{cr}$ and mean annihilation time $\tau_{an}$
may, however, be different, and can be restored in our final results
by a mere replacement $l \to l \tau^*/\tau_{dif}$, $f \to f
\tau^*/\tau_{cr}$ and $g \to g \tau^*/\tau_{an}$.

Note also that the number of particles is not explicitly conserved
in such a dynamical model of the environment due to
 the renewal processes. However,
in the absence of attractive particle-particle interactions and
external perturbations, the particles distribution on the lattice is
uniform and the average occupation $\rho(t) = \overline{\eta({\vec
r})}$ of the lattice tends, as $t \to \infty$, to a constant value,
$\rho_s = f/(f + g)$. This relation
 can be thought of as the Langmuir adsorption
isotherm \cite{langmuir}.

Hence, the limit $\tau_{dif} \to \infty$ (or, $l \to 0$) corresponds
to the ordinary site percolation model with immobile blocked sites.
The limit $f,g \to 0$, ($\tau_{cr}, \tau_{an} \to \infty$), while
keeping the ratio $f/g$  fixed, $f/g = \rho_s/(1-\rho_s)$,
corresponds to the usual hard-core lattice-gas with the conserved
particles number.

At time $t = 0$ we introduce at the  origin of the lattice the
charge carrier, whose role is to probe the response of the
environment modeled by dynamic percolation to an external
perturbation. We stipulate that  only the carrier out of all
participating particles
 can not disappear from the system, and moreover,
its motion is biased by some external constant force. As a physical
realization, we envisage that the carrier is charged, while all
other particles are neutral,
 and the system is exposed to constant external electric
field ${\vec E}$.

The dynamics of the carrier particle is defined as follows: We
suppose  that the waiting time between successive jumps of the
carrier has also an exponential distribution with a mean value
$\tau$, which may  in general be different from the corresponding
waiting time of the environment particles. Attempting to hop, the
carrier first chooses a hop direction with  probabilities
\begin{equation}
\label{trans} p_\mu=exp\Big[\frac{\beta}{2}\Big({\vec E } \cdot
{\vec e}_{
\mu}\Big)\Big]/\sum_{\nu}exp\Big[\frac{\beta}{2}\Big({\vec E} \cdot
{\vec e}_{\nu}\Big)\Big],
\end{equation}
where $\beta$ is the reciprocal temperature,
 ${\vec e}_{\nu}$ (or ${\vec e}_{\mu}$) stand for six unit lattice vectors,
 $\nu,\mu = \{\pm1,\pm2,\pm 3\}$,
connecting the carrier position with $6$ neighboring lattice sites,
and $({\vec E} \cdot {\vec e}_{\nu})$ denotes the scalar product.
 We adopt the
convention that $\pm 1$ corresponds to
 $\pm X$, $ \pm 2 $ corresponds to $\pm Y$ while $\pm 3$ stands for $\pm Z$.
The jump is actually fulfilled when the target lattice site is
vacant. Otherwise, the carrier remains at its position. For
simplicity we  assume in what follows that the external field is
oriented along the $X$-axis in the positive direction, such that
${\vec E} = (E,0,0)$. Note also that for the choice of the
transition probabilities as in Eq.(\ref{trans}), the detailed
balance is naturally preserved.

\section{EVOLUTION EQUATIONS.}

We proceed by writing the evolution equations describing the
dynamics of the system. Let $P({\vec R_{c}},\eta;t)$ denote the
joint probability that at time moment $t$ the charge carrier
occupies position ${\vec R_{c}}$ and all "environment" particles are
in configuration $\eta\equiv\{\eta({\vec r})\}$. Next, let
$\eta^{{\vec r},\mu}$ denote particles' configuration obtained from
$\eta$ by exchanging the occupation variables of the sites ${\vec
r}$ and ${\vec r}+{\vec e}_{\vec \mu}$, i.e. $\eta({\vec
r})\leftrightarrow \eta({\vec r}+{\vec e}_{\vec \mu})$, and
$\hat{\eta}^{{\vec r}}$ be the configuration obtained from $\eta$ by
changing the occupation of the site ${\vec r}$ as $\eta({\vec
r})\leftrightarrow1-\eta({\vec r})$. Clearly, the first type of
process appears due to random hops of the "environment" particles,
while the second one stems from the renewal processes, i.e. random
creation and annihilation of the "environment" particles. Then,
summing up all possible events which can result in the configuration
 $({\vec R_{c}},\eta)$ or change this configuration for any other, we find that the
temporal evolution of the system under study is governed by the
following master equation:
\begin{eqnarray}
&&\partial_tP({\vec R_{c}},\eta;t)=
\frac{l}{6\tau^*}\sum_{\mu}\;\sum_{{\vec r}\neq{\vec R_{c}}-{\vec
e}_{\vec \mu},{\vec R_{c}}}
 \; \left\{ P({\vec R_{c}},\eta^{{\vec r},\mu};t)-P({\vec R_{c}},\eta;t)\right\} +\nonumber\\
&+&\frac{1}{\tau}\sum_{\mu}p_\mu\left\{\left(1-\eta({\vec
R_{c}})\right)P({\vec R_{c}}-{\vec e}_{\vec \mu},\eta;t)
-\left(1-\eta({\vec R_{c}}+{\vec e}_{\vec \mu})\right)P({\vec R_{c}},\eta;t)\right\}+\nonumber\\
&+&\frac{g}{\tau^*}\sum_{{\vec r}\neq {\vec R_{c}}}
\;\left\{\left(1-\eta({\vec r})\right)P({\vec R_{c}},\hat{\eta}^{{\vec r}};t)-\eta({\vec r})P({\vec R_{c}},\eta;t)\right\}+\nonumber\\
&+&\frac{f}{\tau^*}\sum_{{\vec r}\neq{\vec R_{c}}}
\;\left\{\eta({\vec r})P({\vec R_{c}},\hat{\eta}^{{\vec r}};t)
-\left(1-\eta({\vec r})\right)P({\vec R_{c}},\eta;t)\right\}.
\label{eqmaitresse}
\end{eqnarray}
Note that the terms in the first (resp. second) line of
Eq.(\ref{eqmaitresse}) describe random hopping motion of the
"environment" particles (resp. biased motion of the carrier) in
terms of the Kawasaki-type particle-vacancy exchanges, while the
terms in the third and the fourth lines account for the Glauber-type
decay and creation of the "environment" particles.

\subsection{Mean velocity of the charge carrier and correlation functions.}

The velocity of the charge carrier can be now readily determined
from Eq.(\ref{eqmaitresse}). To do it, we multiply both sides of
Eq.(\ref{eqmaitresse}) by $({\vec R_{c}} \cdot {\vec e_1})$ and sum
over all possible configurations  $({\vec R_{c}},\eta)$. This yields
the following expression for the carrier mean velocity $V_{c}(t)$:
\begin{equation}
V_{c}(t)=\frac{\sigma}{\tau}\left\{p_1(1-k({\vec
e_1};t))-p_{-1}(1-k({\vec e_{-1}};t))\right\}, \label{vitesse}
\end{equation}
where $k({\vec \lambda};t)$ denotes the carrier-"environment"
particles pair correlation function:
\begin{equation}
k({\vec \lambda};t)\equiv\sum_{{\vec R_{c}},\eta}\eta({\vec
R_{c}}+{\vec \lambda}) P({\vec R_{c}},\eta;t). \label{defk}
\end{equation}
Note that  $k({\vec \lambda};t)$ is just
 the density distribution of the "environment"
particles, as seen by someone residing on the carrier which moves
with velocity $V_{c}(t)$.

Note now that $V_{c}(t)$ depends explicitly on the local density of
the "environment"
 particles in the immediate vicinity of
the carrier.  Note also that if the "environment" is perfectly
homogeneous, i.e., $k({\vec \lambda};t) \equiv \rho_s$, one has that
$\eta({\vec R_{c}}+{\vec \lambda})$ and $P({\vec R_{c}},\eta;t)$ in
Eq.(\ref{defk}) decouple, which entails, in virtue of from
Eq.(\ref{vitesse}), a trivial mean-field-type result
\begin{equation}
V_{c}^{(0)}=(p_1-p_{-1})(1-\rho_s)\frac{\sigma}{\tau}. \label{vmf}
\end{equation}
The latter (trivial) expression implies that for the perfectly
homogeneous, well-stirred environment the frequency of jumps of the
carrier particles ($\tau^{-1}$) merely gets renormalized by a factor
$1-\rho_s$, which gives the fraction of successful jumps.

The situation, however, appears to be more complex and there emerge
essential backflow effects: As a matter of fact, the carrier
effectively perturbs the spatial distribution of the "environment"
particles so that stationary density profiles emerge. This can be
contrasted to the  earlier dynamic
 percolation models
\cite{drugerA,drugerB,zwanzig,sahimi,chatterjee,granekB,garcia,hilferA,granekS}
in which the carrier had no impact on the embedding medium and hence
there was no re-arrangement of the host medium around the carrier
particle. As a  consequence, in our model   $k({\vec \lambda};t)
\neq \rho_s$,  and $k({\vec \lambda};t)$  approaches
 $\rho_s$ only at infinite separations from the carrier, i.e. when
$|{\vec \lambda}| \to\infty$.  Therefore, we rewrite
Eq.(\ref{vitesse}) in the form
\begin{equation}
V_{c}(t)=V_{c}^{(0)}-\frac{\sigma}{\tau}\{p_1(k({\vec
e_1};t)-\rho_s)-p_{-1}(\rho_s-k({\vec e_{-1};t}))\},
\label{gencorrneg}
\end{equation}
which shows
 explicitly the deviation of the  mean velocity of the carrier from the  mean-field-type result in
Eq.(\ref{vmf}) due to the formation of the density profiles.

\subsection{Evolution equations of the pair correlation functions.}

Equation (\ref{vitesse}) signifies that in order to obtain
$V_{c}(t)$ we have to evaluate
 $k({\vec e_{\pm1}};t)$. Multiplying both sides of Eq.(\ref{eqmaitresse}) by
 $\eta({\vec R_{c}})$ and summing over all configurations  $({\vec R_{c}},\eta)$, we find
  that $k({\vec \lambda};t)$ obeys
\begin{eqnarray}
\partial_tk({\vec \lambda};t)&=& \frac{l}{6 \tau^*} \sum_{\mu}(\nabla_\mu-
\delta_{{\vec \lambda},{\vec e}_{\mu}}\nabla_{-\mu})k({\vec
\lambda};t)-\frac{(f+g)}{\tau^*}k({\vec \lambda};t)+\frac{f}{\tau^*}+\nonumber\\
&+&\frac{1}{\tau}\sum_{\mu} \sum_{{\vec R_{c}},\eta}p_\mu\left(
1-\eta({\vec R_{c}+e}_{ \mu})\right)\nabla_\mu\eta({\vec
R_{c}}+{\vec \lambda})P({\vec R_{c}},\eta;t), \label{evolk}
\end{eqnarray}
where  $\nabla_\mu$ denotes the ascending finite difference operator
of the form
\begin{equation}
\nabla_\mu f({\vec \lambda}) \equiv f({\vec \lambda}+{\vec e}_{
\mu})-f({\vec \lambda}), \label{nabla}
\end{equation}
and
\begin{equation}
\delta_{{\vec r},{\vec r'}} = \left\{\begin{array}{ll}
1,     \mbox{ if the site ${\vec r}={\vec r'}$} \nonumber\\
0,     \mbox{ otherwise.}
\end{array}
\right.
\end{equation}
The  Kroneker-delta term $\delta_{{\vec \lambda},{\vec e}_{\mu}}$
signifies that the evolution of the pair correlations,
Eq.(\ref{evolk}), proceeds differently at large separations and in
the immediate vicinity of the carrier. This stems
 from the asymmetric hopping rules of the carrier particle defined by Eq.(\ref{trans}).

Note next that the contribution in the second line in
Eq.(\ref{evolk}), which stems out of the bias acting on the carrier,
is non-linear with respect to the occupation numbers. In
consequence, the pair correlation function is effectively coupled to
the evolution of the third-order correlations. That is,
Eq.(\ref{evolk}) is not closed with respect to the pair correlations
but rather represents a first equation in the infinite hierarchy of
coupled equations for higher-order correlation functions. One faces,
therefore, the problem of solving an infinite hierarchy of coupled
differential equations and needs to resort to an approximate closure
scheme.

\subsection{Decoupling Approximation.}

Here we employ a simple non-trivial closure approximation, based on
the decoupling of the third-order correlation functions into the
product of pair correlations. More precisely, we assume that for
 ${\vec \lambda}\neq{\vec e}_{\nu}$, the third-order correlation fulfil
\begin{equation}
\label{decouplage} \sum_{{\vec R_{c}},\eta}\eta({\vec R_{c}}+{\vec
\lambda})\eta({\vec R_{c}}+{\vec e}_{ \nu})P({\vec R_{c}},\eta;t)
   \approx k({\vec \lambda};t)k({\vec e}_{\nu};t)
\end{equation}
The approximate closure in Eq.(\ref{decouplage})  has been already
used for studying related models of biased carrier diffusion in
hard-core lattice gases and has been shown to provide quite an
accurate description of both the dynamical and stationary-state
behavior. The decoupling in Eq.(\ref{decouplage})
 was first introduced in Ref.\cite{burlatsky}
 to determine the properties of a driven
carrier diffusion in a one-dimensional hard-core lattice gas with a
conserved number of particles, i.e.  without an exchange of
particles with the reservoir. Extensive numerical simulations
performed in Ref.\cite{burlatsky} have demonstrated that such a
decoupling is quite a plausible approximation for the model under
study. Moreover,  rigorous probabilistic analysis of Ref.\cite{olla}
has shown that for this model the results
 based on the  decoupling scheme in Eq.(\ref{decouplage})
are exact. Furthermore, the same closure procedure has been recently
applied  to study spreading of a hard-core lattice gas from a
reservoir attached to one of the lattice sites \cite{spreading}.
Again, a very good agreement between the analytical results and the
numerical data has been found. Next, the decoupling in
Eq.(\ref{decouplage}) has been used in a recent analysis of a biased
carrier dynamics in
 a one-dimensional model of an adsorbed monolayer in contact
with a vapor phase \cite{benichou}, i.e. a one-dimensional version
of the model to be studied here. Also in this case an excellent
agreement has been observed between the analytical predictions  and
the Monte Carlo simulations data \cite{benichou}. We now show  that
the approximate closure of the hierarchy of the evolution equations
  in Eq.(\ref{decouplage}) allows us to reproduce
 in the limit $f,g=0$ and $f/g = const$
the results of Refs.\cite{nakazato} and \cite{elliott}, which are
known (see e.g. Ref.\cite{kehr}) to provide a very good
approximation for the carrier diffusion coefficient in
three-dimensional hard-core lattice gases with arbitrary particle
density.
 We expect therefore that such a closure scheme will render a plausible
description of the carrier dynamics in our three-dimensional
generalized dynamic percolation model. We base our further analysis
on this approximation.

Making use of Eq.(\ref{decouplage}), we find  from Eq.(\ref{evolk})
that the time evolution of pair correlations is governed by the
following equations. For all ${\vec \lambda}$ except for ${\vec
\lambda} = \{{\bf 0},{\vec e}_{\pm 1},{\vec e}_{\pm 2},{\vec e}_{\pm
3}\}$, we have that $k({\vec \lambda};t)$ obeys
\begin{equation}
 \partial_tk({\vec \lambda};t)=\frac{l}{6 \tau^*}
\tilde{L}k({\vec \lambda};t)+\frac{f}{\tau^*}, \label{systemek1}
\end{equation}
 where
the operator $\tilde{L}$ and coefficients $A_\nu(t)$ are given
explicitly by
\begin{equation}
  \tilde{L}\equiv\sum_\mu A_\mu(t)\nabla_\mu-\frac{6(f+g)}{l},
  \end{equation}
and
\begin{equation}
A_\mu(t) \equiv1+\frac{6\tau^*}{l\tau}p_\mu(1-k({\vec e}_{ \mu};t)),
\label{def}
\end{equation}
the operator $\nabla_\mu$  being defined previously in
Eq.(\ref{nabla}), $\mu=\{\pm1,\pm 2,\pm 3\}$. On the other hand, at
the sites adjacent to the carrier one has
\begin{equation}
\partial_tk({\vec e}_{ \nu};t)=\frac{l}{6\tau^*} \Big(\tilde{L}+
A_\nu(t)\Big)k({\vec e}_{ \nu};t)+\frac{f}{\tau^*},
\label{systemek2}
\end{equation}
where $\nu=\{\pm1,\pm 2,\pm 3\}$.

Now, several comments about equations (\ref{systemek1}) and
(\ref{systemek2}) are in order. First of all, let us note that
Eq.(\ref{systemek2}) represents, from the mathematical point of
view, the boundary conditions for the general evolution equation
(\ref{systemek1}), imposed on the sites in the immediate vicinity of
the carrier. Equations (\ref{systemek1}) and (\ref{systemek2}) have
a different form since in the immediate vicinity of the carrier its
asymmetric hopping rules perturb essentially the "environment"
particles dynamics. Equations (\ref{systemek1}) and
(\ref{systemek2}) possess some intrinsic symmetries and hence the
number of independent parameters can be reduced. Namely, reversing
the field, i.e. changing $ E\to -E$,  leads to the mere replacement
of $k({\vec e_1};t)$ by $k({\vec e_{-1}};t)$ but does not affect
$k({\vec e}_{ \nu};t)$ with $\nu = \{\pm 2, \pm 3\}$, which implies
that
  \begin{equation}
k({\vec e_1};t)(-E)=k({\vec e_{-1};t})(E),\;\;\mbox{and}\;\;k({\vec
e}_{ \nu};t)(-E)=k({\vec e}_{\nu};t)(E) \;\; \mbox{for}\;\; \nu =
\{\pm 2, \pm3\}.
  \label{generalparite}
  \end{equation}
Besides, since  the transition probabilities in Eq.(\ref{trans})
obey $p_2=p_{-2}=p_3=p_{-3}$, one evidently has that $k({\vec
e_2};t)=k({\vec e_{-2}};t)=k({\vec e_3};t)=k({\vec e_{-3}};t)$, and
hence,  by symmetry, $A_2(t)=A_{-2}(t)=A_3(t)=A_{-3}(t)$, which
somewhat simplifies equations (\ref{systemek1}) and
(\ref{systemek2}). Lastly, we note that despite the fact that using
the decoupling scheme in Eq.(\ref{decouplage}) we effectively close
the system of equations on the level of the pair correlations, the
solution of Eqs.(\ref{systemek1}) and (\ref{systemek2}) still poses
serious technical difficulties. Namely, these equations are strongly
non-linear with respect to the carrier velocity, which introduces
the gradient term on the rhs of the evolution equations for the pair
correlation, and  depends by itself on the values of the
"environment" particles densities in the immediate vicinity of the
carrier. Below we discuss a solution to this non-linear problem,
focusing on the limit $t \to \infty$.

\section{SOLUTION OF THE DECOUPLED EVOLUTION EQUATIONS IN THE STATIONARY STATE.}

Consider the limit $t\to\infty$  and suppose that the density
profiles and the stationary velocity of the carrier have non-trivial
stationary values: $k({\vec \lambda})\equiv\lim_{t\to\infty}k({\vec
\lambda};t)$,
  $V_{c}\equiv\lim_{t\to\infty}V_{c}(t)$ and $A_\mu = \lim_{t\to\infty}A_\mu(t)$.
As the next step, we define the local deviations of $k({\vec
\lambda})$ from the unperturbed density: $h({\vec \lambda})\equiv
k({\vec \lambda})-\rho_s$. This yields the following system of
equations:
  \begin{equation}
\tilde{L}h({\vec \lambda})=0,
  \label{systemeh1}
  \end{equation}
for ${\vec \lambda} \neq \{{\bf 0},{\vec e}_{\pm 1}, {\vec e}_{\pm
2},{\vec e}_{\pm 3}\}$, while for the special sites adjacent to the
carrier one has
  \begin{equation}
 (\tilde{L}+A_\nu)h({\vec e}_{\nu})+\rho_s(A_\nu-A_{-\nu})=0,
  \label{systemeh2}
  \end{equation}
Equations (\ref{systemeh1}) and (\ref{systemeh2}) determine the
spatial distribution of the deviation from the unperturbed density
$\rho_s$ in the stationary state.  Note also that in virtue of the
symmetry relations $h({\vec e}_{\pm 2}) = h({\vec e}_{\pm 3})$ and
$A_2 = A_{-2} = A_3 = A_{-3}$.

To solve the coupled non-linear
Eqs.(\ref{vitesse}),(\ref{systemeh1}) and (\ref{systemeh2}) we
proceed in the following, standard manner: We first solve these
equations supposing that the carrier stationary velocity is a fixed,
given parameter (and hence, the functions $A_\nu$ entering
Eqs.(\ref{systemeh1}) and (\ref{systemeh2}) are known). In doing so,
we obtain $h(\lambda)$ in the parameterized form $h({\vec
\lambda})=h({\vec \lambda}; A_{\pm 1},A_{2})$.
 Then, choosing particular values ${\vec \lambda}=\{{\vec e}_{\pm 1},{\vec e}_{\pm 2},{\vec e}_{\pm 3}\}$
and making use of the definition of $A_\mu$, we find a system of
three linear equations with three
 unknowns of the form
  \begin{equation}
A_\nu=1+\frac{6\tau^*}{l\tau}p_\nu\Big(1-\rho_s-h({\vec e}_{\nu};
A_{\pm 1},A_{ 2})\Big),
  \end{equation}
where  $\nu=\{\pm1,2\}$, which will allow us to obtain a closure
relation and hence, to define all $A_\nu$ explicitly (and hence, all
$h(\vec e_{\nu}))$.
 Finally, substituting the results into Eq.(\ref{vitesse}), which can be written down in
terms of $A_\nu$ as
  \begin{equation}
  V_{c}=\frac{l\sigma}{6\tau^*}(A_1-A_{-1}),
  \label{vitimp}
  \end{equation}
we arrive at a closed-form equation determining implicitly the
stationary velocity.

\subsection{Density profiles in the dynamic percolative environment.}

The general solution of Eqs.(\ref{systemeh1}) and (\ref{systemeh2})
 can be obtained in a standard fashion by introducing the  following generating
function:
  \begin{equation}
  H(w_1,w_2,w_3)\equiv\sum_{n_1,n_2,n_3}h(\vec{\lambda}) w_1^{n_1}w_2^{n_2} w_3^{n_3},
  \end{equation}
where $n_1$,$n_2$ and $n_3$ are the components of the vector ${\vec
\lambda}$, ${\vec \lambda}
 = {\vec e}_1 n_1 + {\vec e}_2 n_2
 + {\vec e}_3 n_3$.  Multiplying both sides of Eqs.(\ref{systemeh1}) and (\ref{systemeh2})
 by $w_1^{n_1}w_2^{n_2}
w_3^{n_3}$ and performing summation, we find then that
$H(w_1,w_2,w_3)$ is given explicitly by
  \begin{equation}
  H(w_1,w_2,w_3)=- l \frac{\sum_\nu\Big(A_\nu(w_{|\nu|}^{\nu/|\nu|}-1)h({\vec e}_{\nu})+
\rho_s(A_\nu-A_{-\nu})w_{|\nu|}^\nu\Big)}{l
\sum_{\nu}A_\nu(w_{|\nu|}^{-\nu/|\nu|}-1)-6 (f+g)},
  \label{ddimH}
  \end{equation}
an expression which allows us to determine the stationary density
profiles as seen from
 the carrier which moves with a constant
velocity $V_{c}$.

Inverting next the generating function,
 Eq.(\ref{ddimH}), we get, after rather lengthy but straightforward calculations,  the following explicit
result for the local deviation from the unperturbed density:
  \begin{eqnarray}
\label{dev}
  h(\vec{\lambda})=\alpha^{-1}\Big\{\sum_\nu A_\nu h({\vec e}_{\nu})\nabla_{-\nu}
-\rho_s (A_1-A_{-1})(\nabla_1-\nabla_{-1})\Big\} F(\vec{\lambda}),
  \end{eqnarray}
where $F(\vec{\lambda})$ is given by
  \begin{eqnarray}
  F(\vec{\lambda})=\left(\frac{A_{-1}}{A_1}\right)^{n_1/2}
\int_0^\infty e^{-x} {\rm I}_{n_1}\left(2 \frac{\sqrt{A_1
A_{-1}}}{\alpha} x\right) {\rm I}_{n_2}\left(2 \frac{A_2}{\alpha}
x\right) {\rm I}_{n_3}\left(2 \frac{A_2}{\alpha} x\right) {\rm d}x,
  \end{eqnarray}
and
  \begin{eqnarray}
  \alpha=\sum_\nu A_\nu+\frac{ 6 (f+g)}{l}
= A_1 + A_{-1} + 4 A_2 + \frac{ 6 (f+g)}{l}
  \end{eqnarray}
Consequently, the particles density distribution as seen from the
carrier moving with a constant velocity $V_{c}$ obeys
\begin{eqnarray}
\label{distr} k(\vec{\lambda}) = \rho_s + \alpha^{-1}\Big\{\sum_\nu
A_\nu h({\vec e}_{\nu}\nabla_{-\nu} -\rho_s
(A_1-A_{-1})(\nabla_1-\nabla_{-1})\Big\} F(\vec{\lambda}),
\end{eqnarray}
where we have to determine three yet unknown parameters $A_1$,
$A_{-1}$ and $A_2$.

To determine these parameters, we set in Eq.(\ref{dev}) ${\vec
\lambda} = {\vec  e}_1$, ${\vec  \lambda} = {\vec  e}_{-1}$ and
${\vec  \lambda} = {\vec  e}_{2}$, which results in the system of
three closed-form equations determining the unknown functions
$A_{\nu}$, $\nu = \{\pm1,2\}$,
\begin{equation}
\label{A}
 A_\nu=1+\frac{6\tau^*}{l\tau}p_\nu \left\{1-\rho_s
-\rho_s(A_1-A_{-1})\frac{\det\tilde{C}_\nu}{\det\tilde{C}}\right\},
\end{equation}
where $\tilde{C}$ is a square matrix of the third order defined as

\begin{equation}
\label{m} \pmatrix{ A_1\nabla_{-1}F({\vec e}_{1}) - \alpha &
A_{-1}\nabla_{1}F({\vec e}_{1}) &
                  A_{2}\nabla_{-2}F({\vec e}_{1}) \cr
        A_1\nabla_{-1}F({\vec e}_{-1}) & A_{-1}\nabla_{1}F({\vec e}_{-1}) - \alpha & A_{2}\nabla_{-2}F({\vec e}_{-1})\cr
         A_1\nabla_{-1}F({\vec e}_{2})  &A_{-1}\nabla_{1}F({\vec e}_{2}) &
A_{2}\nabla_{-2}F({\vec e}_{2}) - \alpha}
\end{equation}
while $\tilde{C}_\nu$  stands for the matrix obtained from
$\tilde{C}$ by replacing the  $\nu$-th column by a column vector
$\left((\nabla_1-\nabla_{-1})F({\vec e}_{\nu})\right)_\nu$. Equation
(\ref{distr}),
 together with the definition of the
coefficients $A_{\nu}$,  constitutes the first general result of our
analysis defining the density distribution in the percolative
environment under study.

\subsection{General force-velocity relation.}

Substituting  Eq.(\ref{A}) into (\ref{vitimp}), we find that
 the stationary velocity of the carrier particle is defined implicitly as the solution of equation:
\begin{equation}
\label{velo} V_{c}=\frac{\sigma}{\tau}(p_1-p_{-1})(1-\rho_s)
\left\{1+\rho_s\frac{6\tau^*}{l\tau}\frac{p_1 \det\tilde{C}_1 -
p_{-1} \det\tilde{C}_{-1}}{\det\tilde{C}}\right\}^{-1},
\end{equation}
where $\tilde{C}_{1}$ and $\tilde{C}_{-1}$ are the following square
matrices of the third order:

\begin{equation}
\label{U} \tilde{C}_{1} = \pmatrix{(\nabla_1-\nabla_{-1})F({\vec
e}_{1})  & A_{-1}\nabla_{1}F({\vec e}_{1}) & A_{2}\nabla_{-2}F({\vec
e}_{1}) \cr
      (\nabla_1-\nabla_{-1})F({\vec e}_{-1})   & A_{-1}\nabla_{1}F({\vec e}_{-1}) - \alpha & A_{2}\nabla_{-2}F({\vec e}_{-1})\cr
     (\nabla_1-\nabla_{-1})F({\vec e}_{2})    &A_{-1}\nabla_{1}F({\vec e}_{2}) &
A_{2}\nabla_{-2}F({\vec e}_{2}) - \alpha}\\
\end{equation}

and

\begin{equation}
\label{K} \tilde{C}_{-1} = \pmatrix{ A_1\nabla_{-1}F({\vec e}_{1}) -
\alpha &(\nabla_1-\nabla_{-1})F({\vec e}_{1})&
                  A_{2}\nabla_{-2}F({\vec e}_{1}) \cr
        A_1\nabla_{-1}F({\vec e}_{-1}) &(\nabla_1-\nabla_{-1})F({\vec e}_{-1})& A_{2}\nabla_{-2}F({\vec e}_{-1})\cr
         A_1\nabla_{-1}F({\vec e}_{2})  &(\nabla_1-\nabla_{-1})F({\vec e}_{2})&
A_{2}\nabla_{-2}F({\vec e}_{2}) - \alpha}\\.
\end{equation}
Equation (\ref{velo}) represents our second principal result
defining the force-velocity relation in the dynamic percolative
environment for an arbitrary field and  arbitrary rates of the
diffusive and renewal processes.

\section{CARRIER VELOCITY, FRICTION AND DIFFUSION COEFFICIENTS.}

Consider now the case when the applied external field $E$ is small.
Expanding the transition probabilities $p_1$ and $p_{-1}$ in the
Taylor series up to the first order in powers of the external field,
i.e., setting \be p_{\pm 1} = \frac{1}{6} \pm \frac{\sigma \beta
E}{12} +  {\mathcal O}\Big(E^2\Big), \ee we find that $V_{c}$,
Eq.(\ref{vitimp}), follows \be V_{c} \sim \frac{\sigma}{6 \tau}
\Big\{\sigma \beta E (1 - \rho_s) - (h(\vec{e}_1) -
h(\vec{e}_{-1}))\Big\}. \ee On the other hand, Eq.(\ref{dev})
entails that \be h(\vec{e}_1) - h(\vec{e}_{-1}) = \frac{2 \sigma
\rho_s (1 - \rho_s) \tau^{*}}{l \tau \Big(\alpha_0 {\cal L}(2
A_0/\alpha_0) - A_0\Big) + 2 \rho_s \tau} \beta E  +  {\mathcal
O}\Big(E^2\Big), \ee where \be A_0 = \lim_{E \to 0} A_\nu = 1 +
\frac{\tau^{*}}{l \tau} (1 - \rho_s), \ee and \be \alpha_0 = \lim_{E
\to 0} \alpha = 6 \Big(1 + \frac{\tau^{*}(1 - \rho_s)}{l \tau} +
\frac{f+g}{l}\Big), \ee while
\begin{eqnarray}
{\cal L}(x)\equiv\left\{\int_0^\infty e^{-t}{\rm
I}_0^{2}(xt)\left({\rm I}_0(xt)-{\rm I}_2(xt)\right){\rm
d}t\right\}^{-1}
  =\left\{P({\bf 0};3x)-P(2{\vec e_1};3x)\right\}^{-1},
  \end{eqnarray}
with $P({\vec r};\xi)$ being the generating function,
  \begin{equation}
  P({\vec r};\xi)\equiv\sum_{j=0}^{+\infty}P_j({\vec r})\xi^j,
  \end{equation}
of the probability $P_j({\vec r})$ that a walker starting at the
origin and performing a Polya random walk on the sites of a
three-dimensional cubic lattice will arrive on the $j$-th step to
the site with the lattice vector  ${\vec r}$ \cite{3}.

Consequently, we find that in the limit  of a small applied field $
E$ the force-velocity relation in Eq.(\ref{velo}) attains the
physically meaningful form of the Stokes formula $E = \zeta V_c$,
 which signifies that the frictional force exerted on the carrier
by the environment particles is  \textit{viscous}. The effective
friction coefficient $\zeta$ is the sum of two terms,
\begin{equation}
\zeta=\zeta_0 + \zeta_{coop} \label{fricd}
\end{equation}
where the first term represents a mean-field-type result $\zeta_0
 = 6 \tau/\beta \sigma^2 (1 - \rho_s)$ (see Eq.(7)), while the second one, $\zeta_{coop}$, obeys
\begin{equation}
\zeta_{coop} =  \frac{12 \rho_s \tau^*}{\beta \sigma^2 l (1-\rho_s).
\Big(\alpha_0{\cal L}(2A_0/ \alpha_0)-A_0\Big)}
\end{equation}
The second contribution has a more complicated origin and is
associated with the cooperative behavior - formation of a
inhomogeneous stationary particle distribution around the carrier
moving with constant velocity $V_{c}$. Needless to say, such an
effect can not be observed within the framework of previous models
of dynamic percolation, since there the carrier does not influence
the host medium dynamics
\cite{drugerA,drugerB,zwanzig,sahimi,chatterjee,granekB,garcia,hilferA,granekS}.

Let us now compare the relative importance of two contributions,
i.e. $\zeta_0$ and $\zeta_{coop}$, to the overall friction.
Straightforward analysis shows that the cooperative behavior
dominates at small and moderate $f$ (which entails also small values
of $g$), while for larger $f$, when $\zeta/\zeta_0$ tends to $1$,
 the mean-field behavior
becomes most important. The cooperative behavior also appears to be
more pronounced at larger densities $\rho_s$.

Consider next some analytical estimates. We start with the
situation, in which diffusion of the environment particles is
suppressed, i.e. when $l = 0$. In this case, we get
\begin{equation}
\frac{\zeta_{coop}}{\zeta_0} = \frac{2 \rho_s}{(1 -
\rho_s)\Big(\frac{2}{y} {\cal L}(y) - 1\Big)},
\end{equation}
where
\begin{equation}
y = \frac{1}{3} \Big(1 + \frac{\tau}{\tau^*}\frac{(f + g)}{(1 -
\rho_s)}\Big)^{-1}.
\end{equation}
Suppose first that $\rho_s$ is small, $\rho_s \ll 1$. Then, $y
\approx 1/3(1+\tau/\tau^*(f+g))$ and we can distinguish between two
situations: when $\tau  \ll (f+g)/\tau^*$, i.e. when the carrier
moves faster than the environment re-organizes itself, and the
opposite limit, $\tau  \gg (f+g)/\tau^*$, when the environment
changes  very rapidly compared to the motion of the carrier. In the
former case we find that $y \approx 1/3$, which yields
$\zeta_{coop}/\zeta_0 \approx 2 \rho_s/(6 {\cal L}(1/3) - 1)$,
${\cal L}(1/3) \approx 0.7942$, while in the latter case we have $y
\approx \tau^*/3 \tau (f +g)$ and $\zeta_{coop}/\zeta_0 \approx
\rho_s \tau^*/3 \tau (f + g)$. Note, that in both cases the ratio
$\zeta_{coop}/\zeta_0$ appears to be small, which signifies that at
small densities $\rho_s$ the mean-field friction dominates. Such a
result  is not counterintuitive, of course, since in the absence of
the particles' diffusion, which couples effectively the density
evolution at different lattice sites, no significant cooperative
behavior can emerge at small densities. On the other hand, at
relatively high densities $\rho_s \sim 1$ and $\tau/(1 - \rho_s) \gg
\tau^*/(f +g) \gg \tau$, when the carrier moves at much faster rate
than the host medium reorganizes itself, we find that
$\zeta_{coop}/\zeta_0 \approx \tau^*/3 \tau (f + g) \gg 1$. This
result stems from the circumstance that in sufficiently dense
environments modeled by dynamic percolation a highly inhomogeneous
density profile emerges even in the absence of particles diffusion.
Here, on the one hand, the carrier perturbs significantly the
particle density in its immediate vicinity. On the other hand, the
density perturbation created by the carrier does not shift the
global balance between creation and annihilation events, i.e. the
mean particle density still equals $\rho_s$. The latter constraint
induces then appearance of essential correlations in particles
distribution and hence, appearance of cooperative behavior.

Let us consider the opposite case when the renewal processes are not
allowed, which means that the particles number is conserved and
local density in the percolative environment evolves only due to
particles diffusion. In this case we find
\begin{equation}
\frac{\zeta_{coop}}{\zeta_0} = \frac{2 \tau^* \rho_s}{(l \tau +
\tau^* (1- \rho_s)) \Big(6 {\cal L}(1/3) - 1\Big)}
\end{equation}
Here, the ratio  $\zeta_{coop}/\zeta_0$ can be large and the
"cooperative" friction dominates the mean-field one
 when $l \tau \ll \tau^* (3 \rho_s -1)$, which happens, namely, at
  sufficiently high densities and in the limit when the carrier
moves at a much faster rate than the environment reorganizes itself.
Otherwise, the mean-field friction prevails.

To estimate the carrier particle diffusion coefficient $D_c$ we
assume the validity of the Einstein relation, i.e. $\beta D_{c} =
\zeta^{-1}$ (see, e.g., Ref.\cite{ein}). We find that, in the
general case,  the carrier diffusion coefficient $D_{c}$ reads
\begin{equation}
  D_{c}=\frac{\sigma^2(1-\rho_s)}{6\tau}\left\{1-\frac{2\rho_s\tau^*}{l\tau}\Big(\alpha_0{\cal
L}(2A_0/\alpha_0)-1+\frac{\tau^*
(3\rho_s-1)}{l\tau}\Big)^{-1}\right\}
  \end{equation}
In the particular case of  conserved particles number, when $f,g \to
0$ but their ratio $f/g$ is kept fixed, $f/g = \rho_s/(1-\rho_s)$,
the latter equation reduces to the classical result
  \begin{equation}
\label{nk}
  D_{c}^{NK}=\frac{\sigma^2(1-\rho_s)}{6\tau}\left\{1-\frac{2\rho_s\tau^*}{l\tau} \Big(6 A_0{\cal
L}(1/3)-1+\frac{\tau^*  (3\rho_s-1)}{l\tau}\Big)^{-1}\right\},
  \end{equation}
obtained earlier in Refs.\cite{nakazato} and \cite{elliott} by
different analytical techniques. The result in Eq.(\ref{nk}) is
known to be exact in the limits $\rho_s \ll 1$ and $\rho_s \sim 1$,
and serves as a very good approximation for the self-diffusion
coefficient in hard-core lattice gases of arbitrary density
\cite{kehr}.

Finally,  in the absence of particle diffusion (fluctuating-site
percolation), our result for the carrier particle diffusion
coefficient reduces to
\begin{equation}
  D_{c}^{per}=\frac{\sigma^2(1-\rho_s)}{6\tau}\left\{1-2\rho_s \Big(4[(1-\rho_s)+(f+g)\tau/\tau^*]{\cal
L}(y)+3\rho_s-1\Big)^{-1}\right\}
  \end{equation}
Note, however, that this result only applies when both $f$ and $g$
are larger than zero, such that the renewal processes take place. In
fact, the underlying decoupling scheme is only plausible in this
case. Similarly to the approximate theories in Refs.\cite{nakazato}
and \cite{elliott}, our approach predicts that in the absence of the
renewal processes $D_c^{per}$ vanishes only when $\rho_s \to 1$,
which is an incorrect behavior.

\section{ASYMPTOTIC BEHAVIOR OF THE DENSITY PROFILES.}

The density profiles at large separations in front of and past the
carrier can be readily deduced from the asymptotical behavior of the
following generating function
\begin{equation}
  N(w_1)\equiv\sum_{n_1=-\infty}^{+\infty}h(n_1,n_2=0,n_3=0) w_1^{n_1}.
  \end{equation}
Inversion of Eq.(\ref{ddimH}) with respect to the symmetric
coordinates $n_2$ and $n_3$ yields then
\begin{eqnarray}
&&N(w_1)= \frac{\Big(A_1h(\vec
e_1)+\rho_s(A_1-A_{-1})\Big)\Big(w_1-1\Big) +
 \Big(A_{-1}h(\vec e_{-1})-\rho_s(A_1-A_{-1})\Big)\Big(w_1^{-1}-1\Big)  }{\alpha - A_1 w_1^{-1} - A_{-1} w_1} \times
\nonumber\\
&\times& \int_0^\infty \exp[- x ] {\rm I}_0^{2}(\frac{2 A_2 }{\alpha
-A_1w_1^{-1} -A_{-1}w_1}  x){\rm d}x +
 \frac{4 A_2 h({\vec e}_2 )}{ \alpha - A_1 w_1^{-1} - A_{-1} w_1}  \times \nonumber  \\
&\times&  \int_0^\infty   \exp[-  x]  {\rm I}_0(  \frac{2 A_2
}{\alpha -A_1w_1^{-1} -A_{-1}w_1}   x) \Big({\rm I}_1( \frac{2 A_2
}{\alpha -A_1w_1^{-1}
-A_{-1}w_1} x)- \nonumber\\
&-& {\rm I}_0(\frac{2 A_2 }{\alpha -A_1w_1^{-1} -A_{-1}w_1}
x)\Big){\rm d}x
\end{eqnarray}
We notice now that $N(w_1)$ is a holomorphic function in the region
${\cal W}_1 < w_1 < {\cal W}_2$, where
\begin{eqnarray}
{\cal W}_1 = \frac{\alpha - 4  A_2}{2 A_{-1}} - \sqrt{ \Big(
\frac{\alpha - 4  A_2}{2 A_{-1}}\Big)^{2} - \frac{A_1}{A_{-1}} }
\end{eqnarray}
and
\begin{eqnarray}
{\cal W}_2 = \frac{\alpha - 4  A_2}{2 A_{-1}} + \sqrt{ \Big(
\frac{\alpha - 4  A_2}{2 A_{-1}}\Big)^{2} - \frac{A_1}{A_{-1}} }
\end{eqnarray}
As a consequence, the asymptotic behavior of $h(n_1,n_2=0,n_3=0)$ in
the limit $n_1 \to \infty$ (resp. $n_1 \to - \infty$) is controlled
by the behavior of $N(w_1)$  in the vicinity of $w_1 =  {\cal W}_2$
(resp.  $w_1 =  {\cal W}_1$) (see, for example, the analysis of the
generating function singularities  developed in
Ref.\cite{flajolet}).

\subsection{Asymptotics of the density profiles at large
separations in front of the carrier.}

Consider first the asymptotic behavior of the density distribution
of the "environment" particles at large separations in front of the
carrier. We find then that in the limit $w_1 \to {\cal W}_2$, the
function $N(w_1)$ follows
\begin{eqnarray}
\label{l}
 N(w_1) &\sim&_{w_1 \to {\cal W}_2} \Big[\frac{\Big(A_1h(\vec e_1)+\rho_s(A_1-A_{-1})\Big)\Big({\cal W}_2-1\Big)}{4
\pi A_2 } +
\nonumber\\
&+&\frac{\Big(A_{-1}h(\vec e_{-1})-\rho_s(A_1-A_{-1})\Big)\Big({\cal
W}_2^{-1}-1\Big)}{4 \pi A_2}\Big] \ln\Big({\cal W}_2-w_1\Big)
\end{eqnarray}
Then, (cf, Ref.\cite{flajolet}), we obtain the following
asymptotical result
\begin{equation}
h(n_1,0,0)\sim_{n_1 \to \infty }  \frac{K^+}{n_1}e^{-n_1/
\lambda_+},
\end{equation}
where the characteristic length $\lambda_+$ is given explicitly by
\begin{equation} \lambda_+\equiv
\ln^{-1}\left(\frac{\alpha/2-2A_2}{A_{-1}}+\sqrt{\left(\frac{\alpha/2-2A_2}{A_{-1}}\right)^2-\frac{A_1}{A_{-1}}}\right),
\end{equation} and the amplitude $K^+$ obeys
\begin{eqnarray}
K^+ &=& \Big[\frac{\Big(A_1h(\vec
e_1)+\rho_s(A_1-A_{-1})\Big)\Big({\cal W}_2-1\Big)}{4 \pi A_2 } +
\nonumber\\
&+&\frac{\Big(A_{-1}h(\vec e_{-1})-\rho_s(A_1-A_{-1}\Big)\Big({\cal
W}_2^{-1}-1\Big)}{4 \pi A_2}\Big] > 0,
\end{eqnarray}
which signifies that the density of the "environment" particles in
front of the carrier is higher than the average value $\rho_s$ and
approaches $\rho_s$ at large separations from the carrier as an
\textit{exponential} function of the distance.

\subsection{Asymptotics of the density profiles at large
separations behind the carrier.}

We consider next  the asymptotic behavior of the
 "environment" particles density profiles past the carrier particle, which turns
out to be very different depending on whether the  dynamics of
 the percolative environment obeys the strict conservation of
the "environment" particles number or not (the renewal processes are
suppressed or allowed).

\subsubsection{Non-conserved particles number.}

In the case when  particles  may disappear and re-appear on the
lattice, one has that the root ${\cal W}_1 < 1$. We find then,
following essentially the same lines as in the previous subsection,
that
\begin{eqnarray}
\label{k}
 N(w_1) &\sim&_{w_1 \to {\cal W}_1} \Big[\frac{\Big(A_1h(\vec e_1)+\rho_s(A_1-A_{-1})\Big)\Big({\cal W}_1-1\Big)}{4
\pi A_2 } +
\nonumber\\
&+&\frac{\Big(A_{-1}h(\vec e_{-1})-\rho_s(A_1-A_{-1})\Big)\Big({\cal
W}_1^{-1}-1\Big)}{4 \pi A_2}\Big] \ln\Big(\frac{1}{w_1 -{\cal
W}_1}\Big).
\end{eqnarray}
Hence, in the non-conserved case  the approach to the unperturbed
value $\rho_s$ is also exponential when  $n_1 \to -\infty$, and
follows
 \begin{equation}
  h_{n_1,0,0}\sim_{n_1 \to -\infty}  \frac{K^-}{|n_1|}e^{-|n_1|/ \lambda_-},
\end{equation}
where \be \lambda_-\equiv\ln^{
-1}\left(\frac{\alpha/2-2A_2}{A_{-1}}-\sqrt{\left(\frac{\alpha/2-2A_2}{A_{-1}}\right)^2-\frac{A_1}{A_{-1}}}\right)
\ee and
\begin{eqnarray}
K^- &=& \Big[\frac{\Big(A_1h(\vec
e_1)+\rho_s(A_1-A_{-1})\Big)\Big({\cal W}_1-1\Big)}{4 \pi A_2 } +
\nonumber\\
&+&\frac{\Big(A_{-1}h(\vec e_{-1})-\rho_s(A_1-A_{-1})\Big)\Big({\cal
W}_1^{-1}-1\Big)}{4 \pi A_2}\Big] < 0
\end{eqnarray}
which implies that the particles density past the carrier is lower
than the average. Note that, in the general case,  $\lambda_+ <
\lambda_-$, which means that the depleted region past the carrier is
more extended in space than
 the  traffic-jam-like region in front of the carrier.
The density profiles are therefore asymmetric with respect to the
origin, $n_1 = 0$.

\subsubsection{Conserved particles number.}

Finally, we turn to the analysis of the shape of the density
profiles of the
 percolative environment behind the carrier in the particular limit when the
host medium evolves only due to diffusion, while creation and
annihilation of particles are completely suppressed. In this case,
in which the particles number is explicitly conserved, one has that
for arbitrary value of the field and particles' average density, the
root ${\cal W}_1 \equiv 1$ and, consequently, the form of the
generating function is qualitatively different from that in
Eqs.(\ref{l}) and (\ref{k}),
\begin{eqnarray}
\label{y}
 N(w_1) \sim_{w_1 \to 1^+} \Big[\frac{\Big(A_1h(\vec e_1)-A_{-1}h(\vec
e_{-1})}{4 \pi A_2} + \frac{2\rho_s(A_1-A_{-1})\Big)}{4 \pi A_2
}\Big] (w_1 -1) \ln\Big(\frac{1}{w_1 -1}\Big).
\end{eqnarray}
Equation (\ref{y}) implies that in the limit when the particle
number is conserved the large-$n_1$ asymptotic behavior of
$h_{n_1,0,0}$
 is described by an \textit{algebraic} function of $n_1$ with a logarithmic correction; that is,
\begin{equation}
  h_{n_1,0,0}\sim  \frac{K_- \ln(|n_1|)}{n_1^2},
  \end{equation}
where $K_-$ is an $n_1$-independent constant.  Remarkably, the
power-law decay of correlations
 implies existence of a quasi-long-range order in the percolative environment past the carrier.
 In the conserved case
the mixing of the three-dimensional percolative environment is not
very efficient and there are considerable memory effects - the host
medium  remembers the passage of the carrier on  large space and
time scales.

\section{CONCLUSIONS}

To conclude,  we have presented a microscopic model describing the
dynamics of a charge carrier, driven by an external field ${\vec E}$
in a three-dimensional complex medium modeled by dynamic
percolation, i. e. represented as a cubic lattice partially filled
with mobile, hard-core "environment" particles which can
spontaneously disappear and reappear (renewal processes) in the
system
 with some prescribed rates. Our analytical description of the transfer process in such a medium has been based on
the master equation, describing the time
 evolution of the system,
which has allowed us to evaluate a system of coupled dynamical
equations for the charge carrier velocity and a hierarchy of
correlation functions. To solve these coupled equations, we have
invoked an approximate closure scheme based on the decomposition of
the third-order correlation functions into a product of pairwise
correlations, which has been first introduced in
Ref.\cite{burlatsky} for  a related
 model of a driven carrier dynamics in a one-dimensional lattice gas
with conserved particles number. Within the framework of this
approximation, we have derived a system of coupled, discrete-space
equations describing evolution of the density profiles of the
environment, as seen from the  moving charge carrier, and its
velocity $V_{c}$. We have shown  that
 $V_{c}$ depends on  the density of the "environment" particles  in front of and past the carrier.
Both densities depend on the
 magnitude of the velocity,
as well as on the rate of the renewal and diffusive processes. As a
consequence of such a non-linear coupling, in the general case,
(i.e. for an arbitrary driving field and arbitrary rates of renewal
and diffusive processes), $V_{c}$ has been found only implicitly, as
the solution of a  non-linear equation relating its value to the
system parameters. This equation, which defines the force-velocity
relation for the dynamic percolation under study, simplifies
considerably in the limit of small applied field ${\vec E}$. We find
that in this limit it attains the physically meaningful form of the
Stokes formula, which implies, in particular, that the frictional
force exerted on the carrier by the  environment modeled by dynamic
percolation is \textit{viscous}. In this limit, the carrier velocity
and the friction coefficient are calculated explicitly. In addition,
we determined the self-diffusion coefficient of the carrier in the
absence of the field and show that it reduces to the well-know
result of Refs.\cite{nakazato} and \cite{elliott} in the limit when
the particles number is conserved. Further more, we have found that
the density profile  around the carrier becomes strongly
inhomogeneous: the local density of the "environment"
 particles in front of the carrier is higher than the
average and approaches the average value as an exponential
 function of the distance from the carrier.
On the other hand, behind the carrier the local density is lower
than the average, and depending on whether the number of particles
is explicitly conserved or not, the local density past the carrier
 may tend to the average value either as an exponential or even as an
 \textit{algebraic} function of the distance. The latter reveals
especially strong memory effects and strong correlations between the
particle distribution in the environment and the carrier position.

\section{Acknowledgments.}

G.O.  acknowledges the financial support from the Alexander von
Humboldt Foundation via the Bessel Research Award.

\end{document}